\documentclass[conference]{IEEEtran}

\usepackage[bookmarks=false]{hyperref} 
\usepackage{graphicx} 
\usepackage{balance} 
\usepackage{verbatim} 
\usepackage{fancyhdr} 
\usepackage{amsfonts}
\usepackage{amsmath}
\usepackage{enumitem} 
\usepackage{subfig}
\usepackage{color,soul}

\makeatletter
\def\ps@IEEEtitlepagestyle{%
	\def\@oddfoot{\mycopyrightnotice}%
	\def\@evenfoot{}%
}
\def\mycopyrightnotice{%
	\gdef\mycopyrightnotice{}
}

\ifCLASSINFOpdf
\else
\fi
\begin{document}

%
\title{Management of Solar Energy in Microgrids \\Using IoT-Based Dependable Control}




%
\IEEEoverridecommandlockouts
\author{\IEEEauthorblockN{Manh Duong Phung,
Michel De La Villefromoy, and
Quang Ha}
\IEEEauthorblockA{University of Technology Sydney, 15 Broadway Ultimo NSW 2007, Australia}

Email: manhduong.phung@uts.edu.au
}



\maketitle

\begin{abstract}
Solar energy generation requires efficient monitoring and management in moving towards technologies for net-zero energy buildings. This paper presents a dependable control system based on the Internet of Things (IoT) to control and manage the energy flow of renewable energy collected by solar panels within a microgrid. Data for optimal control include not only measurements from local sensors but also meteorological information retrieved in real-time from online sources. For system fault tolerance across the whole distributed control system featuring multiple controllers, dependable controllers are developed to control and optimise the tracking performance of photovoltaic arrays to maximally capture solar radiation and maintain system resilience and reliability in real time despite failures of one or more redundant controllers due to a problem with communication, hardware or cybersecurity. Experimental results have been obtained to evaluate the validity of the proposed approach.
\end{abstract}

\begin{IEEEkeywords}
Solar tracking, solar energy, dependable control, Internet of things.
\end{IEEEkeywords}

%
\IEEEpeerreviewmaketitle

\section{Introduction}
Solar energy harvesters are becoming essential in home and buildings to provide substantial benefits for our climate, health and economy. It together with other renewable sources will soon outdate the traditional fossil energy to become a pillar for sustainable development. The power harvested by solar arrays, however, depends upon various factors such as weather conditions, photovoltaic (PV) panels, and energy conversion that need extensive and continuous investigations to maximise its potential. To this goal, a number of studies have been conducted using either electromechanical fixtures or customised electronic circuits \cite{Singh2013}.

In the first direction, fixed, single and double axis trackers are developed to track the direction of the sun. In \cite{Roth2004}, an electromechanical system to follow the position of the sun was designed. It is directed by a closed loop servo system that takes measurements of direct solar radiation as the feedback. The control strategy is to drive two small dc motors so that the sun image is kept at the centre of the four-quadrant photo-detector sensing the sun position. Under cloudy conditions, an offline algorithm is used to estimate the position of the sun and takes control of the movement until the detector can sense the sun again. The proposed tracker is simple and cheap, but its accuracy greatly depends on the accuracy of local photo sensors which can be saturated under high temperature and strong radiation conditions. A hybrid tracking system was introduced in \cite{Rubio2007}. It combines open loop tracking strategies based on solar movement models and closed loop strategies using a dynamic feedback controller. The energy consumed by the motors was taken into account during control. However, a global optimisation was not conducted. In \cite{Abdallah2004}, a two-axis sun tracking system using programmable logic controller (PLC) was developed. The open loop was used to control the motion of the sun tracking surface. The results showed an improvement of 41.34\% compared with the fixed surface. Another two-axis solar tracker was implemented in \cite{Nabulsi2010} in which the astronomical method was deployed to determine the position of the sun. The azimuth and elevation angles of PV panels were continuously updated by using a digital signal processor. The two-axis tracking system was also discussed in \cite{Oo2010}, \cite{Seme2011}, and \cite{Tina2012} in which different techniques were employed such as constructing parabolic reflectors, using probabilistic estimating models, or exploiting stochastic search algorithms. Most of them, however, are still based on local sensory information rather than real-time meteorological and astronomical data for optimisation and control.

In the second direction, electronic circuits called maximum power point trackers (MPPT) are designed to drive the PV source operated at the maximum power point (MPP) under different environmental conditions. Those circuits are essentially dc-dc converters providing the correct amount of current so that the load is always supplied with the maximum possible power generated. In \cite{Hussein1995}, the Perturb and Observe (P\&O) technique was studied. This technique bases on the operating voltage of PV arrays and the power they draw to tune the direction of the operating voltage perturbation. Though being low-cost and simple to implement, the P\&O technique causes the operating point to oscillate around the MPP at steady state giving rise to a waste of energy. This problem is mitigated in \cite{Piegari2010} by an adaptive P\&O algorithm that tunes the perturbation amplitude to the actual operating condition. In another work, the incremental conductance algorithm was proposed to deal with partially shaded conditions (PSC) \cite{Ji2011}. It uses the variations of the input voltage and current to detect the occurrence of PSC. The operating point is then moved  from the local MPP to global MPP based on a predetermined linear function. Tey and Mekhilef improved this algorithm by modulating the duty cycle of the dc-dc converter to accelerate the MPP process \cite{Tey2014}. Other popular approaches include Parasitic Capacitance, Voltage based peak power tracking (VMPPT), Current based peak power tracking (CMPPT), fuzzy logic controller, Neural network, and Ripple correlation control (RCC) \cite{Subudhi2013}, \cite{Hohm2000}, \cite{Kottos2006}, \cite{Esram2006}. They, however, have their own limitations that require further investigation, e.g, fuzzy-based controllers perform well under varying atmospheric conditions, but their effectiveness depends on the expert knowledge; VMPP and CMPP trackers are fast to reach the MPP, but the output power is influenced by load characteristics, environmental factors (insulation and temperature), and the type of tracker used.

On another note, studies in both approaches did not address the reliability and resilience aspect of solar trackers which is greatly important for practical deployment. Specifically, solar trackers are often installed at unattended locations connected to the control station via nondeterministic networks. They are thus vulnerable to various types of deterioration such as time-varying transmission delay, hardware failures and communication interruption. They, on the other hand, is also an undetachable part of a large energy management system consisting numerous devices operating in different clusters of microgrid, smart grid and data networks. Dealing with the reliability control of solar tracker therefore requires a new architecture that can avoid service failures and deliver a justifiably trusted service within the context of large distributed control systems (DCS). 

In this work, we propose a new approach to the problem of management of solar energy in microgrids tackling not only the sun tracking problem but also the reliability and resilience of the system. We first introduce a framework to manage the energy flow by integrating various hardware components in a homogeneous network. The connected hardware devices form an IoT network that based on it, embedded computer boards can be employed to control solar trackers. The dependable control technique is then used to enhance the reliability and self-recovering capability of the system. 

\section{System description}
Figure 1 presents the design of our energy management system. The tracking module includes photovoltaic panels (PV) used for collecting solar radiation and motion mechanisms with two degrees of freedom used to control the azimuth and elevation of the panels. The motion can be controlled by either a built-in or a remote controller. The data communication is conducted via a gateway that converts signals from Modbus Serial to Modbus TCP protocols. The collected energy is rectified and stored in two 50 kWh flow-technology batteries. Via inverters and isolators, the AC output is fed onto a separate building microgrid, using standard 3 Phase 415 VAC, that is connected to certain devices and powerpoints placed throughout the building. 

\begin{figure}[!bt]
	\centering
	\includegraphics[width=3.4in]{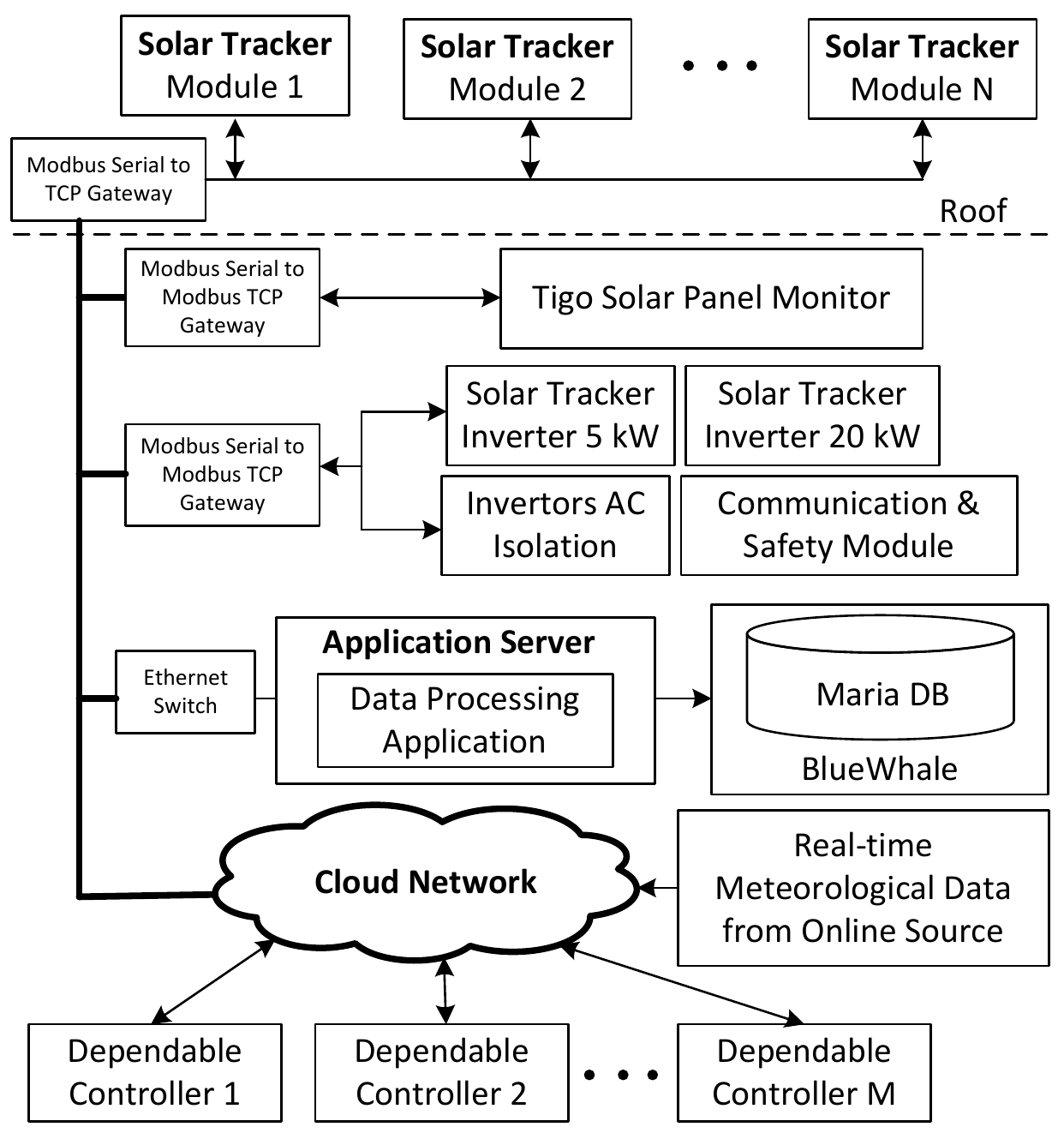}
	\caption{Overview of the solar energy management in microgrids.}
	\label{fig:fig1}
\end{figure}

For monitoring and safety, Tigo systems are implemented to record the solar panel status. The data is then aggregated by algorithms implemented in an application server to analyse the power usage, evaluate the harvested energy and detect abnormal events. The results of this progress, on the one hand, provide references for optimal and control modules in form of network-based application program interfaces (APIs) and, on the other hand, combined with other building data to create energy profiles being stored in databases.

At the core of the system are IoT-based controllers. They are responsible for reliable control of solar trackers under situations of hardware failures and communication interruptions. The controllers are embedded computer boards installed dependable control algorithms. The boards are interconnected and when combining with other processing units form a private cloud network to control and manage the microgrid. The term ``cloud" here implies an abstraction of control hardware introduced to the actuator devices, for instance, a separation between the solar tracker and its controller conducting through standardising signal formats and control protocols provided via network services. Consequently, an actuator device does not need to know which controller is controlling it or how many redundant ones are used for reliability. This abstraction therefore allows one controller to simultaneously control multiple solar trackers and devices while at the same time act as a redundancy for other controllers. It also allows online resources such as real-time meteorological and astronomical data to be retrieved and integrated into the system so that a global optimisation can be conducted resulting in better energy harvested.

\section{Dependable Control}

While the proposed energy management system provides various capabilities and flexibilities for monitoring and managing the solar energy, it requires a control architecture that has the ability to deal with problems relating to cyber- and physical security, hardware failures and data communications. This ability, defined as \textit{dependability} \cite{Avizienis2004}, encompasses reliability, safety, integrity, and availability. While dependability is a concept adopted in the computer systems, dependable control was first introduced just recently \cite{Tran2015}, retaining these attributes by means of feedback control and using the IoT advantages. Its operation bases on the deployment of several controllers to control a single plant, each runs the control algorithm independently. At one time, only one of these controllers is active acting as the duty controller to actually send the control signal to the plant. Status information is concurrently exchanged between controllers during the operation. On top of the status information, performance variables are required to be exchanged between the redundantly backed-up controllers. A standby controller can be activated into the duty role relying on those performance variables from the duty controller. The system is therefore distributed and fault-tolerant. In this section, the design of individual controller is first described. Their cooperation within a dependable framework is then introduced. 

\subsection{Controller Design} \label{sec3a}
Consider a solar tracker having a discrete-time state-space model of the form:
\begin{align}
S: x(k+1) = Ax(k) + Bu(k) \label{Eq1} \\
y(k) = Cx(k),  \label{Eq2}
\end{align}
where $x(k) \in \mathbb{X} \subset \mathbb{R}^n$, $u(k) \in \mathbb{U} \subset \mathbb{R}^m$, and $y(k) \in \mathbb{Y} \subset \mathbb{R}^p$ are the state, control and output vector, respectively. At field level, this system can be controlled by conventional control techniques such as PID or model predictive control (MPC). Here, we consider the state feedback problem having the control $u(k)$ computed online by a MPC algorithm that employs the model (\ref{Eq1}) in prediction, given known state vector $x(k)$. The objective function is defined as follows:
\begin{equation}\label{Eq3}
J(k) =  \sum\limits_{l=1}^N \lVert x(k+l) \lVert^2_Q + \lVert u(k+l-1) \lVert^2_R,
\end{equation}
where $Q$ and $R$ are the weighting matrices and $N$ is the predictive horizon. The minimisation of $J(k)$ subject to (\ref{Eq1}), $x(k) \in \mathbb{X}$, and $u(k) \in \mathbb{U}$ is then solved for the predictive vector sequence defined as $\textbf{u}:=\{u(k), u(k+1),..., u(k+N-1)\}$. The optimal sequence $\textbf{u}^*$ will be obtained as a result of this online computation. However, only the first element $u^*(k)$ is sent to control the solar tracker. Consequently, the optimisation of (\ref{Eq3}) defines an implicit state feedback of the form
\begin{equation}\label{Eq4}
u(k)=K_N(x(k))=u^*(0,x(k)),
\end{equation}
where $u^*(0,x(k))$ is the optimal solution at the time step $k=0,1,...,N-1$ in the horizon, i.e. $u^*(0,x(k))$ is the first vector of the sequence $u^*(k)$.

\begin{figure}[!bt]
	\centering
	\includegraphics[width=3.6in]{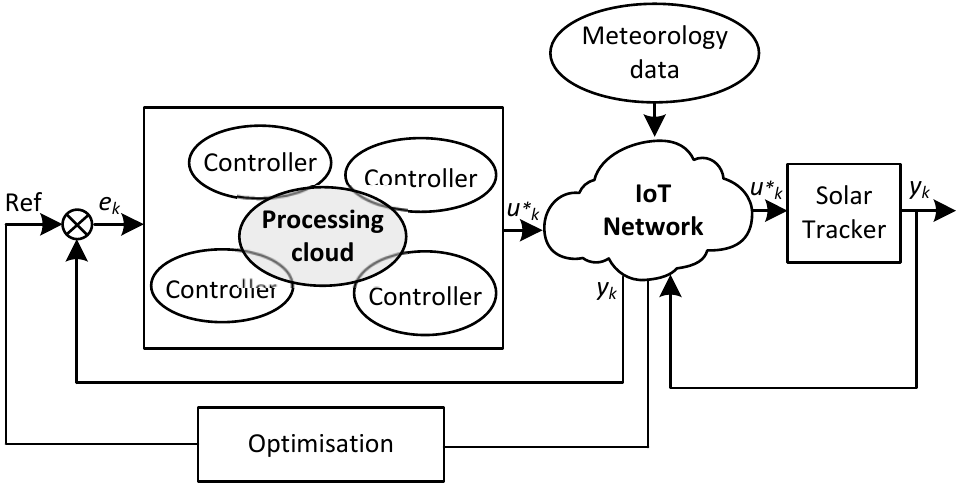}
	\caption{Structure of a dependable control system.}
	\label{fig:fig2}
\end{figure}

\subsection{Dependability for fault-tolerant and self-recovery operation} \label{sec3b}
On top of the low-level control, the dependable control is employed to maintain the system reliability and resilience in case of hardware failures and communication interruptions. It is achieved by using the redundancy together with duty-standby architecture and online switching-over capability. The structure of a dependable control system (DepCS) is presented in \ref{fig:fig2}. A number of \textit{n} processors are designed to involve in a DepCS. The value of \textit{n} depends on the expected integrity level (IL) and available resources. According to industrial data in the computerized-control system field, a structure of four processors would achieve IL-2 of the range 99.99-99.999\%. In a DepCS, each processor will run the control algorithm independently. The set point $ref$ is obtained from the IoT network through an optimisation process. Data for the optimisation such as meteorology information is retrieved from online resources in real time. The generated control signal $u^*(0,x(k))$ and system output $y(k)$ are then transmitted to the plant and controller via the network. Different data transport protocols can be used for communication depending on characteristics of plant and control criteria. For a plant with slow time response, the Transmission Control Protocol (TCP) would be a proper choice as it maximises the reliability. However, if the real time and fast response are more important, the User Datagram Protocol (UDP) and Real-time Transport Protocol (RTP) would be better options. A comparative analysis of transport protocols can be found in \cite{Phung2010}. In this work, we choose to use TCP as the response time of solar trackers is large compared to the data transmission time. 

Besides data communications, the successful implementation of dependable systems rests with the online switching-over capability of controllers. In DepCS, only one controller is active as a duty controller while others are in standby mode. Therefore, the location of the duty processor varies from time to time depending on the state machine. A standby controller can be activated into the duty mode when one of the following events occurs:

\begin{enumerate}[label=(\roman*)]
\item The duty controller is running out of resources requesting a switch to the standby mode;
\item The duty controller is facing hardware failures releasing a switching-over token;
\item The data communication is interrupted triggering a time-out event;
\item An unknown problem occurred causing no state variables to be broadcast.
\end{enumerate}

In these cases, the switching-over process is carried out in a ``first-come, first-served'' manner in which the first standby controller catching the released token will be activated into the duty mode and take the permission to broadcast state variables. 

\begin{figure}[!bt]
	\centering
	\includegraphics[width=3.2in]{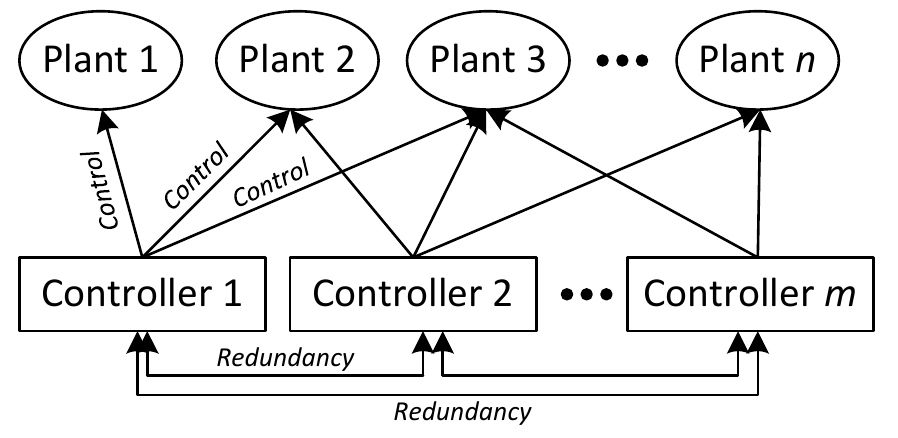}
	\caption{Resource efficiency: one controller controls multiple plants while acting as redundancy for other controllers.}
	\label{figResource}
\end{figure}

\subsection{Resource efficiency in dependable control and Internet of Things}
With IoT, every device is connected to each other and through the cloud infrastructure, the abstraction between them are established. Consequently, no controller is dedicated to a certain device. Instead, one controller can be allocated to simultaneously control multiple plants while, at the same time, act as a backup controller for others as shown in Fig.\ref{figResource}. The number of plants that each controller can control depend on the computation complexity of the control algorithm and the processing capability of the control processor. With rapid advancements in very-large-scale integration (VLSI) design and hardware manufacturing, the embedded computer boards are becoming more powerful with better multi-tasking capability. As a result, a small number of controllers can be deployed to control multiple plants while still maintain the redundancy for reliability and self-recovery.  

\subsection{Stability of dependable control systems}
Given the state feedback control in section \ref{sec3a} and dependable structure in section \ref{sec3b}, constraints on feedback gains are now derived to ensure the stability of DepCS. First, the state and control increment are respectively defined as:
\begin{align}
\Delta{x(k)}:=x(k+1)-x(k) \label{eq6}\\
\Delta{u(k)}:=u(k+1)-u(k) \label{eq7}
\end{align}
with the constraints:
\begin{align}
\lVert \Delta{x(k)}\lVert^2 \le \Delta\alpha \label{eq8}\\
\lVert \Delta{u(k)} \lVert^2 \le \Delta\beta \label{eq9}
\end{align}
for given $\alpha > 0$ and $\beta > 0$. The quadratic supply rate for system $S$ is then defined as:
\begin{equation}
	\psi_{\Delta{k}}:=[\Delta{u(k)}^T \Delta{x(k)}^T]
	\begin{bmatrix}
	P &S \\
	S^T &K
	\end{bmatrix}
	\begin{bmatrix}
		\Delta{u(k)} \\
		\Delta{x(k)}
	\end{bmatrix},
\end{equation}
where $P$, $K$, $S$ are multiplier matrices with symmetric $P$ and $K$. Moreover, $S$ is said to be incrementally dissipative with respect to the supply rate $\psi_{\Delta{k}}$ if there exists a non-negative storage function $V(\Delta{x}):=\Delta{x}^TL_i\Delta{x}, L_i > 0$, such that for all  $\Delta{x}(k)$ and all $k \in Z^+$, the following dissipation inequality is satisfied:
\begin{equation}
	V(\Delta{x}(k+1)) - \tau V(\Delta{x}(k)) \le \psi_{\Delta{k}}, \quad 0< \tau < 1.
\end{equation}
According to \cite{Tran2015}, the closed-loop system $S$ is locally asymptotically stable if $S$ and $u=Kx$ is incrementally dissipative and there are $k_0 \in Z^+$ and $0 < \gamma < 1$ such that the following asymptotic quadratic constraint is fulfilled:
\begin{equation}
	0 < \psi_{\Delta{k}} < \gamma\psi_{\Delta{k}} \quad \forall k \ge k_0. 
\end{equation}

\balance

\begin{figure}[!t]
	\centering
	\subfloat[Solar tracker]{\includegraphics[width=3in]{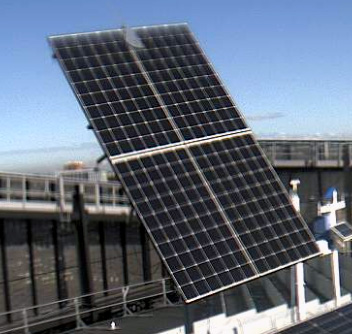}%
		\label{fig3a}}
	\hfil
	\subfloat[PV array]{\includegraphics[width=3in]{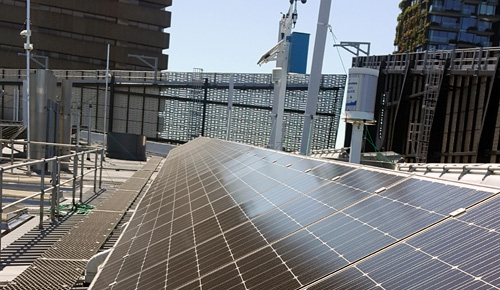}%
		\label{fig3b}}
	\caption{Solar energy harvesters.}
	\label{fig3}
\end{figure}

\begin{figure}[!t]
	\centering
	\subfloat[Inverters and isolators]{\includegraphics[width=1.6in]{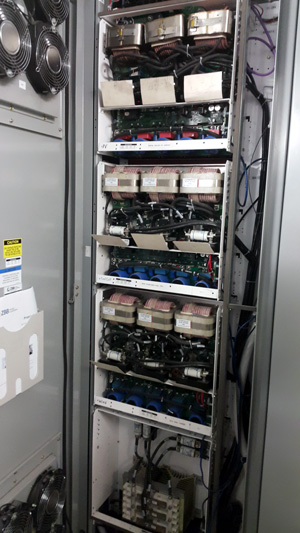}%
		\label{fig4a}}
	\hfil
	\subfloat[Communication gateways]{\includegraphics[width=1.6in]{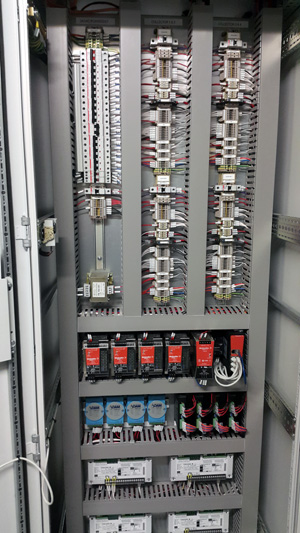}%
		\label{fig4b}}
	\caption{Simulation results for the network.}
	\label{fig4}
\end{figure}

\section{Experimental results}
\begin{figure}[!bt]
	\centering
	\includegraphics[width=3.2in]{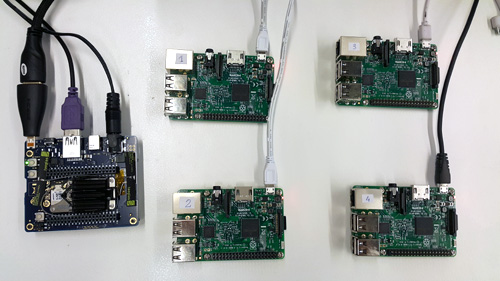}
	\caption{Embedded computer boards used as controllers.}
	\label{fig5}
\end{figure}

Experiments have been conducted to evaluate the validity of the proposed approach. The setup includes a solar tracker and a PV array located at the rooftop of a twelve-storey building as shown in Fig.\ref{fig3}. The solar tracker includes hardware interfaces allowing to control its azimuth and elevation within the range of $0^0$ to $90^0$. The PV array has 72 modules; each has the maximum power of 280 W and the efficiency of 17.2\%. Different inverters and isolators were used to regulate the harvested energy and feed it to the microgrid as shown in Fig.\ref{fig4a}. Fig.\ref{fig4b} shows the communication gateways that convert signals from Modbus Serial to TCP data frames. The IoT boards used to control the solar tracker vary in architectures and configurations, from a quad-core Intel® Atom™ processor 2.4 GHz to a Quad Core 1.2 GHz Broadcom BCM2837 64 bit CPU as shown in Fig.\ref{fig5}. They run Linux operating systems and connect to the network via wireless 802.11 connection. The controllers are executed in the cascade mode in which the inner loop is conducted by the built-in PID controller while the outer loop is managed by the dependable controller. The meteorology information for creating set points is retrieved online with network service APIs and inputs to be 151.1990 in the longitude and -33.8840 in the latitude corresponding to the location of solar trackers. The data communication is carried out via TCP protocol with socket programming.

\begin{figure}[!t]
	\centering
	\subfloat[Normal condition]{\includegraphics[width=3in]{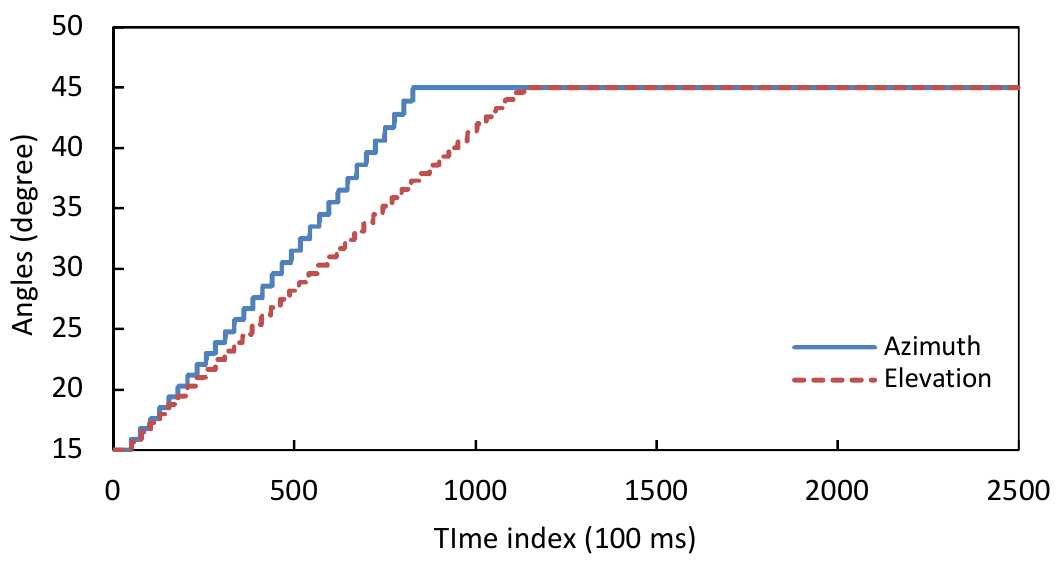}%
		\label{FigResponsea}}
	\hfil
	\subfloat[Subjected to a switching-over event event]{\includegraphics[width=3in]{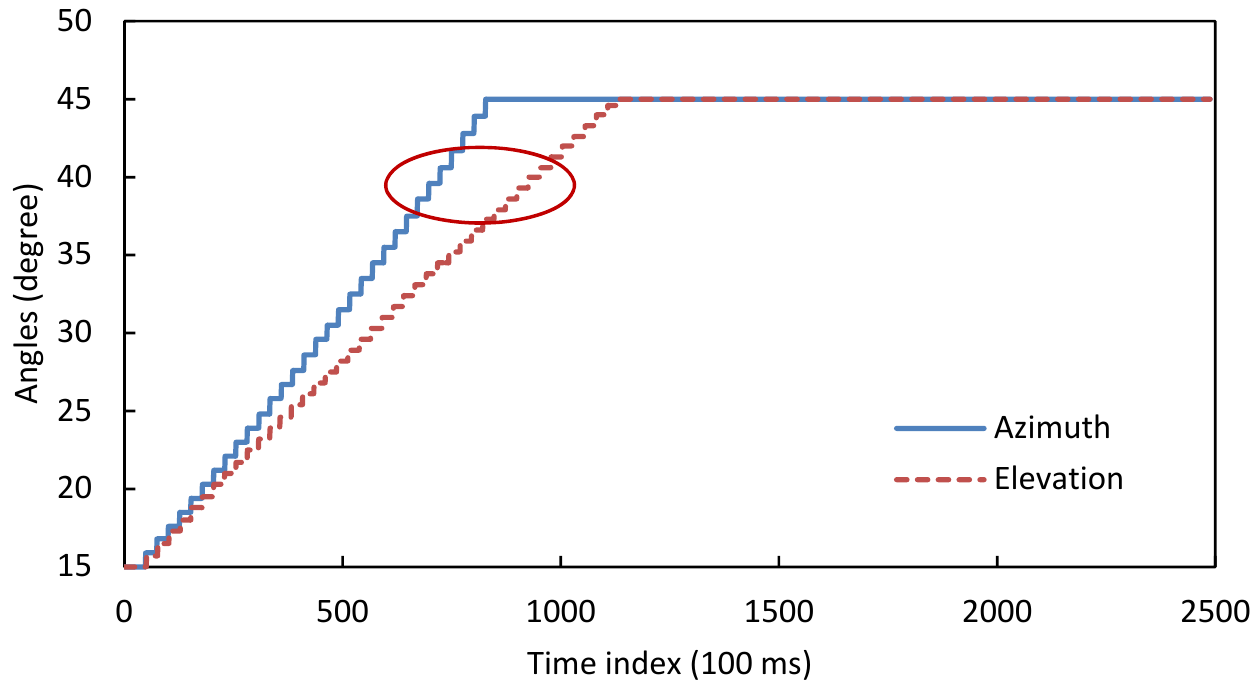}%
		\label{FigResponseb}}
	\caption{Time response of the solar tracker with the azimuth and elevation set to $45^0$.}
	\label{FigResponse}
\end{figure}

\begin{figure}[!bt]
	\centering
	\includegraphics[width=3.2in]{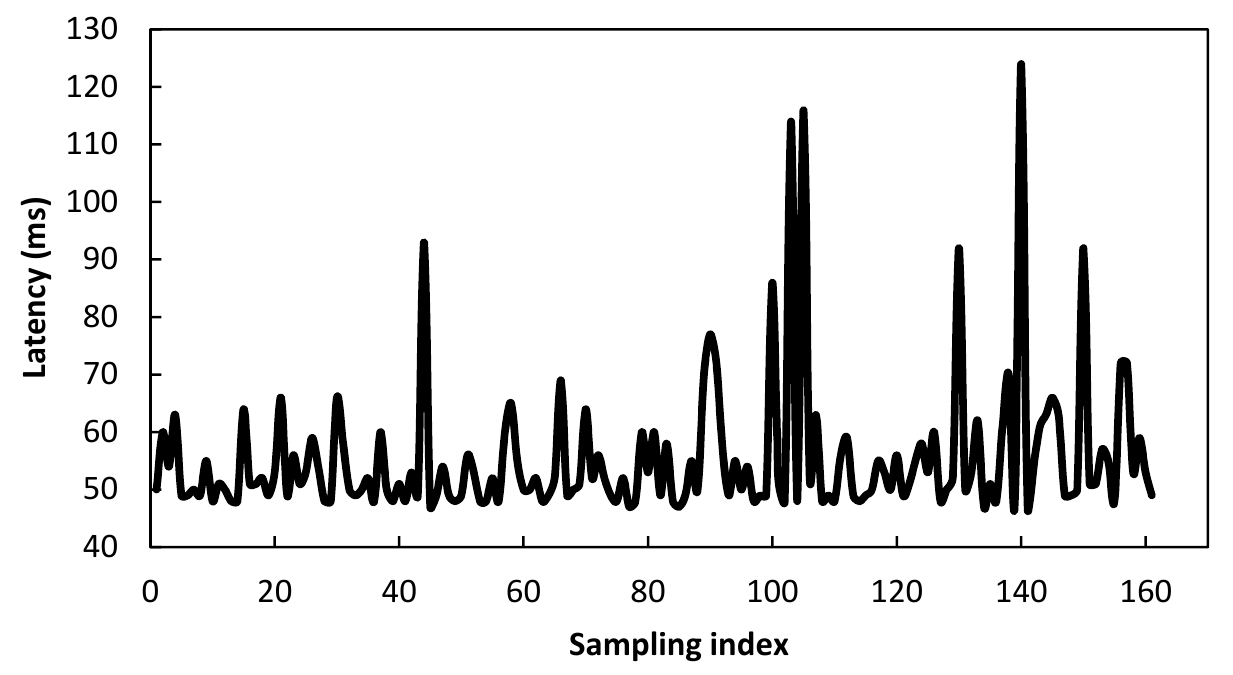}
	\caption{Network latency during the control process.}
	\label{figLatency}
\end{figure}

In experiments, we first test the system capability to track the sun's direction by providing setpoints. Fig.\ref{FigResponsea} shows the output response of the solar tracker with $45^0$ setpoint. Both azimuth and elevation set angles are reached with no overshoots and ripples. The response curves however are rather linear due to configurations of the inner PID loops. The settling time of 100 seconds is slow for motor control, but compared to the change within hours of the sun's direction, this value is obviously enough. Fig.\ref{figLatency} shows the network latency during the control process. The average latency of 55 ms is rather small compared to the control sampling time (100 ms) and settling time (100 s) so that it basically does not affect the control performance. However, at some sampling instances, the latency may be quite large up to 120 ms due to the network jitter causing the control signal to be delayed. In those cases, delay compensation techniques should be considered \cite{Phung2013}, \cite{Hien2009}. 

In next experiments, we evaluated the reliability and self-recovery capability of the system subjected to a hardware failure by intentionally unplugging the power supply of the duty controller during its operation. This action caused the interruption in data communication and led to the occurrence of a time-out event, which is set to 1000 ms, to be broadcast in the system. Without using dependable control, it is impossible to recover the operation of the system after this incident as a regular backup controller could not compute the increment dissipation to maintain the system stability. On the contrary, a standby controller in our dependable system detected the time-out event and switched to the duty mode to take control of the system. It loaded the performance variables recorded from previous communications like setpoints, state variables and sensory outputs to compute the increment dissipation and new control inputs. Fig.\ref{FigResponseb} shows the output response of the system in this situation. The solar tracker eventually reaches the expected operating angles. In fact, with the design using IoT network and dependable control as shown in Fig.\ref{fig:fig2} and Fig.\ref{figResource}, the operation is viable unless the whole system hardware or communication channels are collapsed. Nevertheless, further investigation should be conducted to overcome the downgrade in control performance during the transition process, especially when dealing with fast response systems. 

\begin{figure}[!bt]
	\centering
	\includegraphics[width=3.2in]{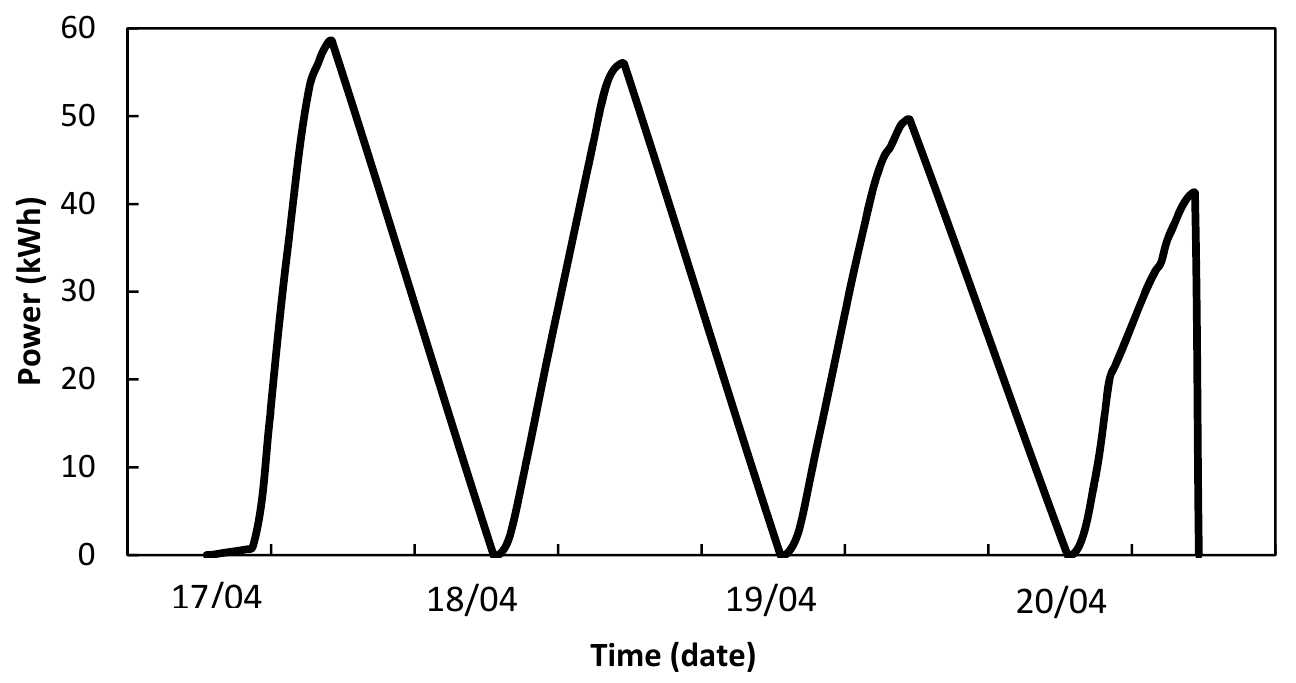}
	\caption{Solar energy generated in 4 days, from 17/04/2017 to 20/04/2017.}
	\label{figEnergy}
\end{figure}

Fig.\ref{figEnergy} shows the real data recording the energy generated by our system within four days. In each day, the energy is accumulated from 6 am to 5 pm. It is able to see that the energy harvested is rather stable between the days with the average value of 48 kWh per day. This amount is sufficient to supply power for various facilities of the local building contributing to its cost reduction, energy diversity and sustainable operation.

\section{Conclusion}
In this work, we have introduced a system to manage the renewable energy harvested by the solar tracker. A hardware and network framework have been successfully implemented with real solar energy fed to the microgrid to power local facilities such as laboratory lighting system and battery charging station. The IoT technology has been utilised to provide ubiquitous computing and control within the microgrid. On top of it, the dependable control technique has been employed to enhance not only the optimal tracking but also the reliability and self-recovery of the system. During the implementation, related concepts including the stability of dependable control systems, transport protocols for real-time data communication, and resource allocation for system efficiency have been discussed providing insights for future energy management systems.  

\section*{Acknowledgement}
This work is supported by the FEIT Blue Sky Research Scheme 2017 of faculty of Engineering and Information Technology, University of Technology Sydney.



%

\end{document}